\documentstyle{article}

\begin{document}

\title{Spectra and generalized eigenfunctions of the\\ one- and two-mode
squeezing operators in quantum optics\thanks {This is a reproduction, with minor corrections and an added Appendix, of a paper published in J. Bertrand et al.(Eds.), {\it Modern Group Theoretical Methods in Physics}, Kluwer Academic Publishers  1995.}}
\author{Bengt Nagel\\
Department of Physics, Division Theoretical Physics\\
Royal Institute of Technology\\
S--100 44 Stockholm  Sweden\\
{\it nagel@theophys.kth.se}}
\date{September 1997}
\maketitle

\section{Introduction and Background Material} \label{S:intro}

\subsection{Introduction and Summary} \label{SS:intrsum}

In recent years the concept of squeezed state has become central in quantum optics both from the theoretical and experimental point of view \cite{JMO87}. In the simplest one-mode case a squeezed state is defined here as a {\it displaced squeezed vacuum} of the form $|\alpha,\zeta \rangle=D(\alpha)S(\zeta)\,|0 \rangle$ obtained by applying the {\it squeezing operator} $S(\zeta)=e^{(\zeta^{*} a^{2}-\zeta a^{\dagger 2})/2}$ \footnote {We follow the convention used e.g. in the first and third references in \cite{JMO87}; sometimes the squeezing operator is defined with the opposite signs in the exponent.} and the {\it displacement operator} $D(\alpha)=e^{
\alpha a^{\dagger}-\alpha^{*} a}$ on the vacuum state $|0\rangle$ (i.e. the ground state of the one-dimensional harmonic oscillator). $\zeta=r\,e^{i2\theta}$ and $\alpha=|\alpha|\,e^{i\beta}$ are complex numbers, and $a=(q+ip)/\sqrt{2},\quad a^{\dagger}=(q-ip)/\sqrt{2}$ in terms of the normalized coordinate and momentum operators $q$ and $p$, where $[q,p]=iI$. The term ``squeezing''comes from the fact that for $cos 2\theta > \tanh r (>0)$ the dispersion $\Delta q$ of the coordinate operator in the squeezed state is smaller than the vacuum state value $1/\sqrt{2}$ : the uncertainty of the value of $q$ is squeezed  compared to the vacuum value. If one performs a rotation in the $qp$-plane (``phase space'') to new canonically conjugate operators $q_{\theta}=\cos \theta\:q+\sin \theta\:p,\; p_{\theta}=-\sin\theta\:q+\cos \theta\:p$ the dispersions $\Delta q_{\theta}$ and $\Delta p_{\theta}$ will be equal to $e^{-r}/\sqrt{2}$ and $e^{r}/\sqrt{2}$, respectively, and the covariance $\Delta(q_{\theta}\, p_{\theta})$ \footnote{We use the notation $\Delta(AB)=\langle(A'B'+B'A')/2\rangle$, where $A'=A-\langle A \rangle$ etc; $\Delta(AA)=(\Delta A)^{2}.$} is zero. This implies that the variance matrix of the original set $q\, p$ (a symmetric $2\times 2$ matrix with the variances $(\Delta q)^{2}$ and $(\Delta p)^{2}$ as diagonal elements and $\Delta(q p)$ as off diagonal elements) has its determinant equal to the minimum allowed value $\frac{1}{4}$. This can actually be taken  as a characteristic property of squeezed states: they are the states (amongst a priori even mixed states) that give equality in the relation $(\Delta q)^{2}\:(\Delta p)^{2}-[\Delta (qp)]^{2} \geq \frac{1}{4}$ (sometimes called the Schr\"{o}dinger--Robertson uncertainty relation).  The coherent states ($\zeta=0$) are characterized by having the variance matrix diagonal with diagonal elements equal to $\frac{1}{2}$. An equivalent characterization of squeezed states uses the fact that $|\alpha,\zeta\rangle$ is the (up to a phase factor unique) normalized vector annihilated by the transformed annihilation operator $b=\cosh r \, (a-\alpha)+e^{i2\theta}\, \sinh r \,(a^{\dagger}-\alpha^{*})$.

The definition of squeezed states we prefer is as the coherent states (with vacuum as isotropy vector) of a representation ({\it rep}) of a group H$\times$ M(2,{\bf R}), which is a semi-direct product of the 3-dimensional Heisenberg-Weyl group H  and the 3-dimensional metaplectic group  M(2,{\bf R}), a two-fold covering of the group SU(1,1).

In experimental realizations of squeezed states the state often corresponds to a two-mode squeezed state generated by applying the two-mode squeezing operator $S_{2}(\zeta)=e^{(\zeta^{*}\,a_{1}\,a_{2}-\zeta\,a_{1}^{\dagger}\,a_{2}^{\dagger})}$ on the vacuum state.

The purpose of this paper is to derive the spectrum and the (generalized) eigenfunctions, in three different forms, of the squeezing generators $(a^{\dagger 2}-a^{2})/4i$ and $(a_{1}^{\dagger}\, a_{2}^{\dagger}-a_{1}\, a_{2})/2i$, and also of another quadratic (in $a$ and $a^{\dagger}$) generator introduced in the next section. Using these results the spectrum and eigenfunctions of the most general second degree polynomial hermitian Hamiltonian (not necessarily bounded below) can be obtained.

The problem of deriving the eigenfunctions of the squeezing generators has been studied before \cite{cL90}, but the results were not complete, in particular the doubling of the spectrum in the one-mode case was not obtained.

In the next section the group theoretical background of the squeezed states will be quickly presented, and after that the three different state representations will be given: the photon number (harmonic oscillator excitation number) representation (the {\it n-representation}). the ordinary configuration space representation (the {\it q-representation}) and the Fock-Bargmann holomorphic function representation (the {\it z-representation}). The last representation has the advantage that the generalized eigenfunctions will be holomorphic functions (but of course not normalizable), whereas in the $q$-representation we expect to get tempered distributions, and in the $n$-representation tempered sequences (growth bounded by some power of $n$). In general the starting point will be the $n$-representation, where the method of an earlier paper \cite{LN70} can be used.

\subsection{Squeezed States as Coherent States of H$\times$M(2,{\bf R})}			   \label{SS:Introsqueezed}

The presentation given here is essentially well-known \cite{aP86} , so we only sketch the derivations. The condition for a linear inhomogeneous transformation of \{$a,\,a^{\dagger}$\}
\begin{equation} 
 b=\mu\,a+\nu\,a^{\dagger}+\alpha,\quad b^{\dagger}=\nu^{*}\,a+\mu^{*}\,a^{\dagger}+\alpha^{*}
 \end{equation}
to give a new set \{$b,\,b^{\dagger}$\} satisfying the canonical commution relation (CCR) is $|\mu|^{2}-|\nu|^{2}=1$, which means that the homogeneous part of the transformation is an element of the group SU(1,1). It is known that any two irreducible representations ({\it irreps}) of the CCR are unitarily equivalent. Now the original rep by \{$a,\,a^{\dagger}$\} is clearly irreducible, and it is not hard to convince oneself that this also holds for the rep by \{$b,\,b^{\dagger}$\} --- one can show that there is a unique ``new vacuum'' $|0\rangle_{b}$ with $b \, |0\rangle_{b}=0$. This means that one gets a set of unitary operators $\{V(\alpha,\,g):\alpha \in {\bf C},\,g \in {\rm SU}(1,1)\}$ , determined up to phase factors, such that 
\begin{equation}
 b=V(\alpha,\,g)^{\dagger}\,a\,V(\alpha,\,g)  
\end{equation}
Composing two successive transformations and using irreducibility one realizes that one has a rep up to a phase factor of the group ISU(1,1), ``inhomogeneous SU(1,1)'', with the composition $(\alpha', g' )\cdot (\alpha, g)=(\alpha' +g' \alpha, g'  g)$, where $g ' \alpha=\mu'  \alpha+\nu' \,\alpha^{*}$. It is then known that the obstruction against choosing the phase factors of $V(\alpha,\,g)$ in such a way that one gets a true rep is present already in the 2-dimensional translation subgroup $\{\alpha\}$: since $a$ and $a^{\dagger}$ don't commute one has to make a (central) extension of this group to the 3-dimensional Heisenberg-Weyl group H$=\{(\alpha,\,s):\alpha \in {\bf C},\,s \in{\bf R}\}$ with the composition $(\alpha',s' )\,(\alpha,s)=(\alpha' +\alpha,s' +s+{\rm Im}[\alpha'  \alpha^{*}])$. Furthermore it turns our that the resulting rep of SU(1,1) is two-valued, so that if one introduces the two-fold covering of SU(1,1), called the metaplectic group M(2,{\bf R}), one obtains a true unitary rep of the group H$\times$ M(2,{\bf R}). Here already the restriction to the group H is irreducible, whereas the restriction to M(2,{\bf R}) gives the sum of two irreps. Corresponding to a factorization of a general group element in the form
\begin{equation}
		(\alpha,g,s)=(\alpha,e,0)\cdot(0,g,0)\cdot(0,e,s)
\end{equation}
we can write a general representing unitary operator in the product form
\begin{equation}
		U(\alpha,g,s)=D(\alpha)U(g)e^{is}.
\end{equation}
The standard Perelomov definition of a set of coherent states of a given rep of a group G starts by choosing a particular isotropy vector of the rep which is invariant (up to a phase factor, i.e. gives a one-dimensional rep) under a suitably chosen subgroup H of the given group. The set of corresponding coherent states is the set of states (ignoring differences in phase factors) obtained by applying all unitary operators in the rep on this vector. The set can then be parametrized by the elements of the homogeneous space G/H.

In our case we choose the vacuum state as isotropy vector (as a more general case one could choose an arbitrary number state $|n\rangle$ to get {\it displaced  squeezed number states}), and we then want to factorize the SU(1,1)-representing operator as
\begin{equation}
 U(g)=S(\zeta)\cdot e^{i2\varphi\,J_{0}}
\end{equation}
where as earlier $S(\zeta)=e^{(\zeta^{*}a^{2}-\zeta a^{\dagger 2})/2}$, and $J_{0}=(a^{\dagger}a+aa^{\dagger})/4=(N+\frac{1}{2})/2$ is the generator of a U(1) subgroup of SU(1,1). As
\begin{equation}
		S(\zeta)^{\dagger}a\,S(\zeta)=\cosh r\cdot a-e^{i2\theta}\sinh r \cdot a^{\dagger},\quad e^{-i2\varphi J_{0}}\,  a\, e^{i2\varphi J_{0}}=e^{i\varphi}\, a,
\end{equation}
we get by comparing with $U(g)^{\dagger}a \, U(g)=\mu  a+\nu a^{\dagger}$ the relations
\begin{equation}
		\mu=e^{i\varphi}\cosh r,\quad \nu=-e^{i(2\theta-\varphi)} \sinh r.
\end{equation}
$J_{0}$ together with $J_{1}=(a^{\dagger 2}+a^{2})/4$ and $J_{2}=(a^{\dagger 2}-a^{2})/4i$ generates a rep
\begin{equation}
[J_{0},J_{1}]=iJ_{2},\quad [J_{0},J_{2}]=-iJ_{1}, \quad [J_{1},J_{2}]=-iJ_{0}
\end{equation}
of the Lie algebra su(1,1) with Casimir operator $C_{2}=J_{0}^{2}-J_{1}^{2}-J_{2}^{2}=-\frac{3}{16}$. We get two irreps of the two-fold covering of SU(1,1) belonging to the discrete series $D_{k}^{+}$ with $k=\frac{1}{4}$ and $\frac{3}{4}$ ($C_{2}=k(k-1)$ in an irrep), corresponding to the spectrum $Sp\,J_{0}=\{\frac{1}{4},\frac{5}{4},\dots\}$ (spanned by the even number states), and $\{\frac{3}{4},\frac{7}{4},\dots\}$ (spanned by the odd number states). We shall also study the generator $K_{+}=J_{0}+J_{1}=(a+a^{\dagger})^{2}/4$. $J_{0},\,J_{2}$, and $K_{+}$ generate reps of the three different types of one-parameter subgroups of SU(1,1): elliptic, hyperbolic, and parabolic, respectively. Since two irreps are involved, from general results it follows that the spectrum of $J_{2}$ is twice the whole real line, and for $K_{+}$ twice the positive half-axis. It remains to give explicit expressions for the corresponding generalized eigenfunctions. As mentioned earlier this will be done in three different forms, which we now proceed to define.

\subsection{Three Different Representations of Oscillator States}
\label{SS:threediff}

The connection between the three different representations introduced earlier, the {\it n-,q-} and {\it z-}representations, can be made via the ON bases correspondences as follows:
\begin{equation}
\{|n\rangle \}\,\leftrightarrow \,\{u_{n}(q)=N_{n}\,H_{n}(q)\cdot e^{-q^{2}/2}\}\,\leftrightarrow \,\{z^{n}/\sqrt{n!} \},\,{\rm where}\, N_{n}=(\sqrt{\pi}\, 2^{n} n!)^{-\frac{1}{2}}
\end{equation}
\begin{equation}
|\varphi\rangle =\sum {a_{n}Ê|n\rangle}\,\leftrightarrow \,\varphi(q)=\sum a_{n}\,u_{n}(q)\, \leftrightarrow \,\phi(z)=\sum a_{n}\,z^{n}/\sqrt{n!}\,=\,e^{|z|^{2}/2}\,\langle z^{*}|\varphi\rangle,
\end{equation}
where $\langle\alpha|\varphi\rangle$ is the so-called coherent state representation;
\begin{equation}
\|\varphi\|^{2}=\sum |a_{n}|^{2}=\int|\varphi(q)|^{2}\,dq\,=\pi^{-1}\,\int \int  |\phi(z)|^{2}e^{-|z|^
{2}}\,d^{2}z.
\end{equation}

The generators of the representation of our group has the forms
\begin{eqnarray}
 &\quad a \leftrightarrow (q+d/dq)/\sqrt{2} \leftrightarrow d/dz, \quad a^{\dagger} \leftrightarrow (q-d/dq)/\sqrt{2} \leftrightarrow z; \nonumber \\
J_{0}:&\quad (a^{\dagger}a+aa^{\dagger})/4 \leftrightarrow (-d^{2}/dq^{2}+q^{2})/4=H_{osc}/2\hbar\omega \leftrightarrow (zd/dz+1/2)/2; \nonumber \\
J_{1}:& \quad (a^{\dagger 2}+a^{2})/4 \leftrightarrow (d^{2}/dq^{2}+q^{2})/4 \leftrightarrow (d^{2}/dz^{2}+z^{2})/4; \\
J_{2}:& \quad (a^{\dagger 2}-a^{2})/4i \leftrightarrow i(qd/dq+1/2)/2 \leftrightarrow i(d^{2}/dz^{2}-z^{2})/4; \nonumber \\
K_{+}:&\quad (a^{\dagger}+a)^{2}/4 \leftrightarrow q^{2}/2 \leftrightarrow (d^{2}/dz^{2}+2zd/dz-z^{2}+1)/4. \nonumber
\end{eqnarray}

We shall need formulas taking us from $\varphi(q)$ to $\phi(z)$ and vice versa (Bargmann and inverse Bargmann transforms). These are easily obtained using the integrals for $a_{n}$:
\begin{equation} 
a_{n}=\int u_{n}(q)\, \varphi (q)\,dq=\pi^{-1}\int \int (z^{*n}/\sqrt{n!}) \phi(z)\,e^{-|z|^{2}}\,d^{2}z
\end{equation}

and the generating function for the Hermite polynomials
\begin{equation} 
\exp(2yt-t^{2})=\sum H_{n}(y)\,t^{n}/n!.
\end{equation}
The result is
\begin{equation} 
\phi(z)=\int K(z,q)\varphi(q)\,dq; \quad \varphi(q)= \pi^{-1}\int\int K(z^{*},q) \phi(z)\,e^{-|z|^{2}}\,d^{2}z,
\end{equation}
where the kernel
\begin{equation} 
K(z,q)=\pi^{-1/4}\, \exp(-z^{2}/2-q^{2}/2+\sqrt{2}\,zq).
\end{equation}

\section{Generalized Eigenfunctions of 
the Squeezing Generator $J_{2}$} \label{S:gener}

We start by deriving the eigenfunctions in the {\it n-}representation. Since the eigenfunctions belong to a continuous spectrum, the sequence $\{a_{n}\}$ we obtain will not be square summable, only power bounded (i.e. a tempered sequence, dual to rapidly decreasing test sequences). Se e.g. \cite{LN70} for details about the mathematical background. So we have to solve the eigenvalue equation
\begin{equation}  
J_{2}|\lambda\rangle =[(a^{\dag2}-a^{2})/4i]|\lambda\rangle =\lambda|\lambda\rangle .
\end{equation}
With $|\lambda\rangle =\sum f_{n}(\lambda)|n\rangle$ this leads to the recursion formula
\begin{equation} 
\sqrt{n(n-1)}f_{n-2}(\lambda)-\sqrt{(n+1)(n+2)}f_{n+2}(\lambda)=i4\lambda f_{n}(\lambda),
\end{equation}
evidently connecting only the even or the odd number states, corresponding to a doubling of the spectrum. To get rid of the square root we put ($N(\lambda)$ is a normalization constant)
\begin{equation} 
f_{n}(\lambda)=N(\lambda)\sqrt{\Gamma([n+1]/2)/\Gamma(n/2+1)}\,a_{n}
\end{equation}
to get the relation
\begin{equation} 
(n+1)a_{n+2}+i4\lambda a_{n}-na_{n-2}=0.
\end{equation}

We first treat the case of even numbers $n$ and put $n=2m$ and $b_{m}=a_{2m}$, so that
\begin{equation}Ê
(m+\frac{1}{2})b_{m+1}+i2\lambda b_{m}-mb_{m-1}=0.
\end{equation}
This recursion relation can be solved by the standard Laplace method introducing the expression $b_{m}=\int t^{m}\,y(t)\,dt$ in the relation, which gives after partial integrations
\begin{equation} 
\int t^{m}[-d(t^{2}y)/dt+ty/2+i2\lambda y+dy/dt]\,dt -{\bf I}\,t^{m}(1-t^{2})y(t)=0
\end{equation}
where {\bf I...} gives the contribution from the ends of the integration path. Taking this path to be from -1 to 1 (so that the corresponding contribution vanishes) and solving the simple first order differential equation for $y(t)$ we get after an integration variable change leading to an integral from 0 to 1, and using the standard integral representation for the hypergeometric function $_{2}F_{1}$ (which for convenience we denote by {\it F}) the result
\begin{equation}
b_{m}=(-1)^{m}2^{-1/2} \frac{\Gamma (1/4-i\lambda) \Gamma(1/4+i\lambda)}{\Gamma(1/2)} F(-m,1/4-i\lambda,1/2;2)
\end{equation}

One could also have found this result directly by comparing the recursion relation with a suitable contiguity relation for the hypergeometric functions \cite{EI53}.\\
A transformation formula for transforming $z\rightarrow z/(z-1)$ in {\it F} gives
\begin{equation}
F(-m,1/4-i\lambda,1/2;2)=(-1)^{m}F(-m,1/4+i\lambda,1/2;2).
\end{equation}
Together with the relation, valid for real $\lambda$,
\begin{equation}
F(-m,1/4-i\lambda,1/2;2)^{*}=F(-m,1/4+i\lambda,1/2;2)
\end{equation}
this implies, using the series expansion of $F$ , that $F$ is for even {\it m} a real, even polynomial of degree {\it m} in $\lambda$, whereas for odd {\it m} it is {\it i} times a real, odd polynomial of degree {\it m}. With a change of definition of the normalization factor we can write 
\begin{equation} 
f_{2m}(\lambda)=N(\lambda)\sqrt{(1/2)_{m}/m!}\,F(-m,1/4+i\lambda,1/2;2),\, m=0,1,..
\end{equation}
using the notation  $(a)_{m}=\Gamma(a+m)/\Gamma(a)$.\\
Treating the odd {\it n} case in a similar way we obtain
\begin{equation}
f_{2m+1}(\lambda)=N\rq (\lambda)\sqrt{(3/2)_{m}/m!}\,F(-m,3/4+i\lambda,3/2;2),\, m=0,1,...
\end{equation}

We now have to prove the (generalized) orthonormality and completeness relations of these generalized eigenfunctions, and in particular determine the normalization constants. Although this can be done using (26) and (27) and suitable integral representations it is simpler to observe that we have obtained special cases of the Pollaczek polynomials \cite{EII53} \footnote {This definition differs with the factor $i^{m}$ from the definition in the original article, and makes $P_{m}$ a real polynomial with positive coefficient for the highest power term}
\begin{equation}
P_{m}(\lambda,b)=i^{m}\sqrt{(2b)_{m}/m!}\,F(-m,b+i\lambda,2b;2),\, m=0,1,..
\end{equation}
which for every $b>0$ form a complete orthonormal set of polynomials on the real line with the weight function \footnote{see Appendix}
\begin{equation}
\rho_{b}(\lambda)=2^{2b-1}|\Gamma(b+i\lambda)|^{2}/\pi\Gamma(2b).
\end{equation}
If we choose positive normalization factors
\begin{equation}
N(\lambda)=\sqrt{\rho_{1/4}(\lambda)},\quad N'(\lambda)=\sqrt{\rho_{3/4}(\lambda)}
\end{equation}
and define
\begin{equation}
|\lambda,e\rangle=\sum f_{2m}(\lambda)|2m\rangle, \quad f_{2m}=(-i)^{m}\sqrt{\rho_{1/4}(\lambda)}\,P_{m}(\lambda,1/4), 
\end{equation}
\begin{equation}
|\lambda,o\rangle=\sum f_{2m+1}(\lambda)|2m+1\rangle,\quad f_{2m+1}=(-i)^{m}\sqrt{\rho_{3/4}(\lambda)}\,P_{m}(\lambda,3/4),
\end{equation}
then the generalized orthonormality relations would be
\begin{equation}
\langle\lambda,e|\lambda',e\rangle=\langle\lambda,o|\lambda' ,o\rangle=\delta(\lambda-\lambda'),\quad\langle\lambda,e|\lambda', o\rangle=0.
\end{equation}
Whereas the last relation is obvious, the first two just express the completeness of the corresponding set of Pollaczek polynomials, and thus hold true.

In a similar way the orthonormality conditions for the Pollaczek polynomials imply the completeness relation for the generalized eigenfunctions, since this relation says that
\begin{equation}
\langle\psi|\psi\rangle=\int[|\langle\lambda,e|\psi\rangle|^{2}+|\langle\lambda,o|\psi\rangle|^{2}]\,d\lambda
\end{equation}
for an arbitrary state $|\psi\rangle=\sum c_{n}|n\rangle$, which can also be written in the form
\begin{equation}
\sum|c_{n}|^{2}=\sum_{m,n}[c_{2m}^{*}c_{2n}\int f_{2m}(\lambda)^{*}f_{2n}(\lambda)\,d\lambda\,+\,c_{2m+1}^{*}c_{2n+1}\int f_{2m+1}(\lambda)^{*}f_{2n+1}(\lambda)\,d\lambda].
\end{equation}
As $P_{0}(\lambda,b)=1$ it follows from (31) and (32) that
\begin{equation}
\langle 0|\lambda,e\rangle=\sqrt{\rho_{1/4}(\lambda)},\quad \langle1|\lambda,o\rangle  = \sqrt{\rho_{3/4}(\lambda)}
\end{equation}

To derive the form of the eigenfunctions in the {\it z}-representation one could of course start from the corresponding eigenvalue equation
\begin{equation}
(-d^{2}/dz^{2}+z^{2})\phi_{\lambda}(z)=i4\lambda\phi_{\lambda}(z).
\end{equation}
We shall instead use the result we have obtained in the {\it n}-representation and apply a standard integral representation for the {\it F} function to obtain from (26)
\[
\phi_{\lambda e}(z)=\sum f_{2m}z^{2m}/\sqrt{(2m)!}=A(\lambda)\int_{0}^{1}t^{-3/4-i\lambda}(1-t)^{-3/4+i\lambda} \left[ \sum \frac{\sqrt{(1/2)_{m}}}{\sqrt{m!(2m)!}} [z(2t-1)]^{m}\right] \] 
\begin{equation}
=A(\lambda)\,e^{-z^{2}/2}\int_{0}^{1}t^{-3/4-i\lambda}(1-t)^{-3/4+i\lambda}e^{z^{2}t}\,dt,\mbox{ where}\, A(\lambda)=\sqrt{\rho_{1/4}(\lambda)} \frac{\Gamma(1/2)} {|\Gamma(1/4+i\lambda)|^{2}}.
\end{equation}
Use of a standard integral representation for the confluent hypergeometric function $_{1}F_{1}$ results in the final form
\begin{equation}
\phi_{\lambda,e}(z)=\sqrt{\rho_{1/4}(\lambda)}\,e^{-z^{2}/2}\,_{1}F_{1}(1/4-i\lambda,1/2;z^{2}).
\end{equation}
In a similar way we obtain
\begin{equation}
\phi_{\lambda,o}(z)=\sqrt{\rho_{3/4}(\lambda)}\,e^{-z^{2}/2}\,z\,_{1}F_{1}(3/4-i\lambda,3/2;z^{2}).
\end{equation}
As the second parameter of the $_{1}F_{1}$ function is 1/2 or 3/2, respectively, the functions can be expressed in term of the parabolic cylinder functions $D_{\nu}(\sqrt{2}\,z)$ and $D_{\nu}(-\sqrt{2}\,z)$ with $\nu=-1/2+i2\lambda$ (see \cite{EII53}, p.117).

Finally we shall derive the form of the eigenfunctions in the {\it q}-representation. In this case we shall solve the corresponding eigenvalue equation, which can be written
\begin{equation}
q\,d\varphi_{\lambda}(q)/dq=-(\frac{1}{2}+i2\lambda)\,\varphi_{\lambda}(q).
\end{equation}
We have a singularity at $q=0$ and can choose independent solutions on the positive and negative half-axes. These solutions can be put together to give solutions even and odd, respectively, in {\it q}, corresponding to even and odd number solutions. We then get
\begin{equation}
\varphi_{\lambda e}(q)=(2\pi)^{-1/2}\frac{1}{|q|^{1/2+i2\lambda}},\quad \varphi_{\lambda o}=(2\pi)^{-1/2}\frac{{\rm sgn}\, q}{|q|^{1/2+i2\lambda}}.
\end{equation}
The factors $(2\pi)^{-1/2}$ are introduced to ensure the generalized orthonormality relation
\[
\int\varphi_{\lambda'}(q)^{*}\,\varphi_{\lambda}(q)\,dq  =(2\pi)^{-1} 2\int_{0}^ {\infty} |q|^{-1+i2(\lambda' -\lambda)}dq 
\]
\begin{equation}
= (2\pi)^{-1} \int_{-\infty}^{\infty} \exp[i(\lambda '-\lambda)t]\,dt=\delta(\lambda'-\lambda).
\end{equation}
It is clear that $\varphi_{\lambda e}(q)$ and $\varphi_{\lambda o}(q)$ should correspond to the earlier determined eigenfunctions up to $\lambda$-dependent phase factors. To determine these we calculate the scalar product with the vacuum and 1-photon state, respectively, and compare with (36):
\begin{equation}
\pi^{-1/4}\int e^{-q^{2}/2}\varphi_{\lambda e}(q)\,dq=\pi^{-1/4} (2/\pi)^{1/2}\int_{0}^{\infty} q^{-1/2-i2\lambda}e^{-q^{2}/2}\,dq=(2\pi^{3})^{-1/4}\frac{\Gamma(1/4-i\lambda)}{2^{i\lambda}},
\end{equation}
\begin{equation}
\pi^{-1/4} \sqrt{2}\, \int q e^{-q^{2}/2} \varphi_{\lambda o}(q)\,dq =2(2\pi^{3})^{-1/4}\frac{\Gamma(3/4-i\lambda)}{2^{i\lambda}}.
\end{equation}
From this we finally get
\begin{eqnarray}
\langle q|\lambda,e\rangle = \exp[i\{\arg\Gamma(1/4+i\lambda)+\lambda\, \ln2\}](2\pi)^{-1/2}\frac{1}{|q|^{1/2+i2\lambda}},\\
\langle q|\lambda,o\rangle=\exp[i\{\arg\Gamma(3/4+i\lambda)+\lambda \,\ln2\}](2\pi)^{-1/2}\frac{{\rm sgn}\,q}{|q|^{1/2+i2\lambda}}.
\end{eqnarray}
It should be observed that although the eigenfunctions in the {\it q}-representation are ordinary functions (though with a singularity at the origin) they are not square integrable but should be considered as tempered distributions on the real line, similar to the eigenfunctions $e^{i\lambda q}$ of the momentum operator $p=-i\,d/dq$. See e.g. \cite{LN70} for more details on the mathematical nature of the generalized eigenfunctions of the SU(1,1) generators.

\section{Generalized Eigenfunctions of the Generator $K_{+}$}
\label{S:genk}

We have seen in (12) that the parabolic generator $K_{+}$ has the forms $(a^{\dagger}+a)^{2}/4$,
$q^{2}/2$, and $(d^{2}/dz^{2}+2z\,d/dz-z^{2}+1)/4$ in the different representations. As in the previous case one could of course start from the eigenvalue  equation in the $n$-representation and solve the corresponding difference equation, which again involves either even or odd number states, corresponding to a doubling of the spectrum. In this case it is simpler to use the form of the operator in the $q$-representation, where the eigenvalue equation
\begin{equation}
(q^{2}/2)\varphi_{\eta}(q)=\eta\,\varphi_{\eta}(q) \mbox{ or } (q^{2}-2\eta)\varphi_{\eta}(q)=0
\end{equation}
shows that the spectrum of $K_{+}$ is the half axis $[0,\infty)$ and that we can choose as properly normalized even and odd solutions
\begin{eqnarray}
\varphi_{\eta e}(q)&=&(1/2\sqrt{2\eta})\,[\delta(q-\sqrt{2\eta})+\delta(q+\sqrt{2\eta})],\\
\varphi_{\eta o}(q)&=&(1/2\sqrt{2\eta})\,[\delta(q-\sqrt{2\eta})-\delta(q+\sqrt{2\eta})],
\end{eqnarray}
where the normalization is determined (up to a $\eta$-dependent phase factor) by demanding
\begin{equation}
\int\varphi_{\eta' e}(q)^{*}\varphi_{\eta e}(q)\,dq=\delta(\eta' -\eta) \mbox{ etc}.
\end{equation}
Completeness of the set of eigenfunctions follows from the fact that together the two families give a family of $\delta$-functions covering the whole $q$-axis.

The corresponding eigenfunctions in the $n$- and $z$-representations are obtained from (13) and (15) as
\begin{eqnarray}
|\eta,e\rangle=\sum g_{2m}(\eta)\,|2m\rangle,\;|\eta,o\rangle=\sum g_{2m+1}(\eta)\,|2m+1\rangle, \mbox{ where}\\ 
g_{2m}(\eta)=\int\varphi_{\eta e}(q)\,u_{2m}(q)\,dq= N_{2m}(1/\sqrt{2\eta})\,e^{-\eta}H_{2m}(\sqrt{2\eta}) \\
g_{2m+1}(\eta)=\int\varphi_{\eta o}(q)\,u_{2m+1}(q)\,dq= N_{2m+1}(1/\sqrt{2\eta})\,e^{-\eta}H_{2m+1}(\sqrt{2\eta}) 
\end{eqnarray}

\begin{eqnarray}
f_{\eta e}(z)=\int K(z,q)\varphi_{\eta e}(q)\,dq=\pi^{-1/4}(1/\sqrt{2\eta})\,e^{-\eta}e^{-z^{2}/2}\cosh(2\sqrt{\eta}\,z),\\
f_{\eta o}(z)=\int K(z,q)\varphi_{\eta o}(q)\,dq=\pi^{-1/4}(1/\sqrt{2\eta})\,e^{-\eta}e^{-z^{2}/2}\sinh(2\sqrt{\eta}\,z).
\end{eqnarray}

\section{Spectrum of the General Second Degree Hamiltonian}
\label{S:spec}

A general hermitian second degree polynomial in $\{a,a^{\dagger}\}$ (or in $\{q,p\}$) can be written
\[
p_{2}(a,a^{\dagger})=A(a^{\dagger}a+a\,a^{\dagger})+Be^{i\Phi}a^{2}+Be^{-i\Phi}a^{\dagger 2}+Ce^{i\Psi}a+Ce^{-i\Psi}a^{\dagger}+D\,I\]
\begin{equation}
=4[AJ_{0}+B(\cos\Phi\,J_{1}+\sin\Phi\,J_{2})]+C\sqrt{2}(\cos\Psi\,q-\sin\Psi\,p)+D\,I.
\end{equation}
Since we have a real linear combination of a set of generators of the group H$\times $M(2,{\bf R}) we can conclude that if we have a Hamiltonian of type (57) with arbitrary time-dependent coefficients $A(t),\:B(t),\:C(t) \mbox{ and } D(t)$, the corresponding time development will be given by a unitary operator of the form (4), and hence a squeezed state will always (up to a phase factor) develop into a squeezed state. This clearly also holds for the set of displaced squeezed number states for a given $n$.

We shall now determine the various possible spectra for the operator (57). It is no essential restriction to assume $A,\:B, \mbox{ and } C$ non-negative, and $D=0$.

If $A\,\neq \,B$ we can transform away the linear terms (in $C$) and get the quadratic terms to a form proportional to $J_{0}$ (if $A\,>\,B$) or to $J_{2}$ (if $A\, < \,B$) by a substitution $a\rightarrow \mu\,a+\nu\,a^{\dagger}+\alpha$, which according to (2) can be achieved by a unitary transformation. We get the results
\begin{eqnarray}
\mbox{If } A\,>\,B :&\quad 4\sqrt{A^{2}-B^{2}}\,J_{0}-[C^{2}/2(A^{2}-B^{2})][A\cos(\Phi-\Psi)-B\cos\Psi];  \\ 
\mbox{If } A\,<\,B :&\quad 4\sqrt{B^{2}-A^{2}}\,J_{2}-[C^{2}/2(A^{2}-B^{2})][A\cos(\Phi-\Psi)-B\cos\Psi].
\end{eqnarray}
In both cases we should choose $\alpha=-C\,[Ae^{-i\Psi}-Be^{i(\Psi-\Phi)}]/[2(A^{2}-B^{2})].$ 

If $A\,=\,B$ the phase rotation $a\rightarrow e^{-i\Phi/2}a$ results in the form
\begin{equation} 
2Aq^{2}+C\sqrt{2}\cos(\Psi-\Phi/2)\,q-C\sqrt{2}\sin(\Psi-\Phi/2)\,p.
\end{equation}
Thus if $C\,=\,0$, or $C\,\neq \,0$ and $\Phi\,=\,2\Psi$ (mod $2\pi$) we get the form
\begin{equation}
2A(q\pm C/2\sqrt{2}\,A)^{2}-C^{2}/4A.
\end{equation}
Otherwise, i.e. if $C\,\neqÊ\,0$ and $\Phi\,\neq \,2\Psi$ (mod $2\pi$) a translation in $q$ leads to the form
\begin{equation}
2A[q^{2}-(C/\sqrt{2}\,A)\sin(\Psi-\Phi/2)\,p]-(C^{2}/4A)\cos^{2}(\Psi-\Phi/2).
\end{equation}
From the relation
\begin{equation}
\exp(iq^{3}/3b)(q^{2}+bp)exp(-iq^{3}/3b)=bp
\end{equation}
it follows that in this case the operator can be transformed to the form
\begin{equation}
-C\sqrt{2}\sin(\Psi-\Phi/2)\,p \quad (\mbox{or } \times q).
\end{equation}
From (58), (59), (61), and (64) we can read off the different possible spectra of $p_{2}(a,a^{\dagger})$ in (57). $A\,=\,B$ is evidently a critical point. Here we can have as spectrum either a half-axis taken twice or the whole real axis. This situation is not stable: a small perturbation can lead either to a discrete equidistant semi-bounded spectrum (if $\,A\,>\,B$) or the whole real line taken twice (if $\,A\,<\,B$).

\section{The Two-Mode Case}
\label{S:thetwo}

In the two-mode case (two-dimensional harmonic oscillator) we get an infinite set of irreducible (true) representations of SU(1,1) by putting
\begin{equation}
J_{0}=(a_{1}^{\dagger}a_{1}+a_{2}^{\dagger}a_{2}+1)/2,\: J_{1}=(a_{1}^{\dagger}a_{2}^{\dagger}+a_{1}a_{2})/2,\: J_{2}=(a_{1}^{\dagger}a_{2}^{\dagger}-a_{1}a_{2})/2i.
\end{equation}
For the Casimir operator we obtain
\begin{equation} 
C_{2}=J_{0}^{2}-J_{1}^{2}-J_{2}^{2}=([\Delta N]^{2}-1)/4,\quad \Delta N=N_{1}-N_{2}=a_{1}^{\dagger}a_{1}-a_{2}^{\dagger}a_{2}.
\end{equation}
Since the spectrum of $\Delta N$ is the set of all integers, we get for the spectrum of $C_{2}=k(k-1)$ as possible $k$ values 1/2, 1, 3/2,.., i.e. all the irreducible SU(1,1) reps in the discrete class $D_{k}^{+}$. All except the first one (with $\Delta n$=0) occur twice. It is evidently enough to study those with $\Delta n\geq 0$. We get as spectrum for $J_{0}\:\{(\Delta n+1)/2+n_{2};n_{2}=0,1,2,..\}$. The generalized eigenfunctions for $J_{2}$ and $K_{+}$ can be derived in the $n$-representation (basis $\{|n_{1},n_{2}\rangle\}$) and from there obtained in the $z$- and $q$-representations (functions $\phi(z_{1},z_{2})$ and $\varphi(q_{1},q_{2})$). We shall only give the results for both operators in the first two representations. The integrals leading to the $q$-representation seem rather intractable in the general case; we only give the result for $J_{2}$ in the case $\Delta n=0$.

The results for the two-mode squeezing generator $J_{2}\:(\Delta n=0,1,..) $ are
\begin{eqnarray} 
|\lambda,\Delta n\rangle=\sum_{0}^{\infty}f_{n}(\lambda,\Delta n)|n+\Delta n, n\rangle,\quad f_{n}(\lambda,\Delta n)=\sqrt{\rho_{c}(\lambda)}\,P_{n}(\lambda,c) \\
\phi_{\lambda}(z_{1},z_{2};\Delta n)=\sqrt{\rho_{c}(\lambda)/\Gamma(2c)}\:z_{1} ^{\Delta n}\,exp(-z_{1}z_{2})\,_{1}F_{1}(c-i\lambda,2c;2z_{1}z_{2}).
\end{eqnarray}
$P_{n}(\lambda,c)$ and $\rho_{c}(\lambda)$ are defined in (28) and (29). Here $c=(\Delta n+1)/2$. \\
For $\Delta n < 0$ one should substitute $|\Delta n|$ for $\Delta n$ in (67) and (68), interchange $z_{1}$ and $z_{2}$, and change $|n+\Delta n,n\rangle$ to $|n,n+|\Delta n|\rangle$.\\
The generalized orthonormality and completeness properties can be written
\begin{eqnarray}
\langle\lambda \rq,\Delta n\rq|\lambda,\Delta n\rangle = \delta_{\Delta n\rq,\Delta n}\:\delta(\lambda\rq-\lambda), \\
\sum_{\Delta n} \int|\lambda,\Delta n\rangle \langle\lambda,\Delta n|\,d\lambda =I,\: \mbox{identity operator}.
\end{eqnarray}.

In the $q$-representation we only give the result for the case $\Delta n=0$:
\begin{equation}
\varphi_{\lambda}(q_{1},q_{2};0)=\frac{1}{\pi\,|\Gamma(1/2+i\lambda)|}\left|\frac{q_{1}-q_{2}}{q_{1}+q_{2}}\right |^{i\lambda}\:K_{i\lambda}(|q_{1}^{2}-q_{2}^{2}|/2),
\end{equation}
where $K_{\nu}$ is the modified Bessel function of the third kind.

For the two-mode generator $K_{+}=J_{0}+J_{1}$ we get
\begin{equation}
|\eta,\Delta n\rangle=\sum g_{n}(\eta,\Delta n)\,|n+\Delta n,n\rangle,\quad g_{n}=(-1)^{n}\sqrt{2}\,e^{-\eta}(2\eta)^{\Delta n/2}\sqrt{\frac{n!}{(n+\Delta n)!}}\,L_{n}^{\Delta n}(2\eta),
\end{equation}
where $L_{n}^{\alpha}$ is the generalized Laguerre polynomial.
\begin{equation} 
\phi_{\eta}(z_{1},z_{2};\Delta n)=\sqrt{2}\,e^{-\eta}(z_{1}/z_{2})^{\Delta n/2}\,e^{-z_{1}z_{2}}I_{\Delta n}(\sqrt{2\eta\,z_{1}z_{2}}),
\end{equation}
where $I_{\nu}$ is the modified Bessel function of the first kind. \\
Changes for $\Delta n <0$, orthonormality and completeness relations are analogous to those in the previous case.

\section{Dedication}
\label{S:D}

This paper is a contribution to professor Guy Rideau on the occasion of his retirement. Rideau has been an important member of the group of scientists in Paris and Dijon that over the years has inspired my interest in group representation theory. May his retirement be as pleasant as mine!

\appendix
\section{Completeness and orthonormality of the set of polynomials $\{P_{n}(\lambda ,b);n=0,1,2..\}$ }

In this Appendix we want to show that the set 
$\{P_{n}(\lambda ,b);n=0,1,2..\},\mbox{ where } P_{n}(\lambda ,b)=  \\ i^{n}\sqrt{(2b)_{n}/n!}\,_{2}F_{1}(-m,b+i\lambda,2b;2) $, 
satisfies the relations
\begin{equation} \label{A1}
\int_{-\infty}^{\infty}P_{m}(\lambda ,b)\,P_{n}(\lambda 
,b)\, \rho_{b}(\lambda)\, d\lambda= \delta_{m n},\quad \mbox{(orthonormality)}
\end{equation}
\begin{equation} \label{A2}
\sum_{n=0}^{\infty}P_{n}(\lambda,b) \, P_{n}(\lambda',b)=\delta(\lambda-\lambda')
/\rho_{b}(\lambda), \quad \mbox{(completeness)} 
\end{equation}
where $\rho_{b}(\lambda)=2^{2b-1}|\Gamma(b+i\lambda)|^{2}/\pi\Gamma(2b)$.

These two relations can be proved using appropriate integral representations for the $_{2}F_{1}$ function; we shall do this, but first make an alternative derivation of the completeness relation, based on the recursion relation satisfied by the $P_{n}$ polynomials. This derivation has the advantage of giving the expression for the weight function $\rho $.

Using the definition of $P_{n}$ and the contiguity relation for $_{2}F_{1}$ referred to in [5] we find the recursion relation generalizing (18)
\begin{equation} \label{A3}
c_{n-1}P_{n-1}+c_{n}P_{n+1}=\lambda\:P_{n};\quad c_{n}=\frac{1}{2}\sqrt{(n+1)(n+2b)};\quad  P_{0}=1.
\end{equation}

Using the notation $P_{n}'=P_{n}(\lambda',b)$ we can derive from (\ref{A3})
\begin{equation} \label{A4}
c_{M}\Delta_{M}=(\lambda-\lambda')\sum_{n=0}^{M}P_{n}\,P_{n}';\quad \Delta_{M}=P_{M}'P_{M+1}-P_{M+1}'P_{M}.
\end{equation}
This is the difference equation analogue of a well-known formula for linear second order differential equations. We shall derive the  leading asymptotic behaviour of $\Delta_{M}$ as $M\rightarrow\infty$. To do this we need the corresponding asymptotic behaviour of the $_{2}F_{1}$ function.

\subsection{Asymptotic behaviour of $_{2}F_{1}(-n,b+i\lambda,2b;2)$ as $n\rightarrow\infty$}

Using a standard integral representation of the $_{2}F_{1}$ function and dividing the integration interval into parts $(0,1/2)$ and $(1/2,1)$ we can write, with $N(\lambda,b)=\Gamma(2b)/\Gamma(b+i\lambda)\Gamma(b-i\lambda)$,
\[ _{2}F_{1}(-n,b+i\lambda,2b;2)=N(\lambda,b) \int_{0}^{1}t^{b+i\lambda-1}(1-t)^{b-i\lambda-1} (1-2t)^{n}\,dt  \]
\begin{equation} \label {A5}
=N(\lambda,b)    \int_{0}^{1/2}[t(1-t)]^{b-1} \left\{ \left (\frac{t}{1-t}\right)^{i\lambda}+(-1)^{n}\left(\frac{t}{1-t}\right)^{-i\lambda}\right\}e^{n \ln(1-2t)}\,dt.
\end{equation}

It is clear from the form of the last integral that the dominant contribution for large $n$ comes from the behaviour near 0. Introducing $x=-n\ln(1-2t)$ as new integration variable and expanding to the first two leading orders in $x/n$ (actually the full asymptotic expansion can in principle be obtained by continuing this expansion), we get after some calculations, using among other things the standard integral representation of the gamma function, the result (we assume $\lambda$ real)
\begin{eqnarray} \label{A6}
_{2}F_{1}(-n,b+i\lambda,2b;2)=&(-i)^{n}[2\Gamma(2b)/|\Gamma(b+i\lambda)|](2n)^{-b}\{\cos(\lambda \ln2n-\varphi-n\pi/2) \nonumber \\ &-(1/n)[b\sqrt{b^{2}+\lambda^{2}}]\cos(\lambda\ln2n-\varphi_{1}-n\pi/2)+O(1/n^{2})\},
\end{eqnarray}
where $\varphi=\arg\Gamma(b+i\lambda),\:\varphi_{1}=\arg\Gamma(b+1+i\lambda)=\varphi+\arctan(\lambda/b)$.

(\ref{A6}) is the result we shall use to derive the asymptotic behaviour of $\Delta_{M}$.

\subsection{Asymptotic behaviour of $\Delta_{M}$ as $M\rightarrow \infty$ and the completeness relation for the set $\{P_{n}\}$}

Using the result (\ref{A6}) in the expression $\Delta_{M}=P_{M}'P_{M+1}-P_{M+1}'P_{M}$ we obtain after some elementary calculations, using also the relation $\Gamma(M+b)/\Gamma(M+1)\approx M^{b-1}$
\begin{equation} \label{A7}
\Delta_{M}=[\Gamma(2b)/2^{2b-2}|\Gamma(b+i\lambda)|^{2}](1/M)\{\sin[(\lambda-\lambda')\ln2M] +(\lambda-\lambda') O(1/M)\}.
\end{equation}

From (\ref{A7}) we finally obtain the completeness relation of the polynomials $\{P_{n}\}$ in the form
\begin{equation} \label{A8}
\sum_{n=0}^{\infty}P_{n}(\lambda,b)\,P_{n}(\lambda',b)=\lim_{M \rightarrow \infty}c_{M}\Delta_{M}/(\lambda-\lambda') = [\pi\Gamma(2b)/2^{2b-1}|\Gamma(b+i\lambda)|^{2}] \,\delta(\lambda-\lambda'),
\end{equation}
where we have used the well-known representation of the $\delta$-function
\begin{equation} \label{A9}
\delta(\lambda-\lambda')=\lim_{n\rightarrow\infty}\frac{\sin n(\lambda-\lambda')}{\pi\,(\lambda-\lambda')}.
\end{equation}

\subsection{Completeness of the set $\{P_{n}\}$ using an integral representation}

Using the integral representation of the $_{2}F_{1}$ function in the first line of (\ref{A5}) and the fact that $P_{n}$ is real, $P_{n}(\lambda,b)^{*}=P_{n}(\lambda,b)$, if $\lambda$ real, we can write, inverting summation and integration,
\begin{eqnarray} \label{A10}
\sum_{n=0}^{\infty} P_{n}(\lambda,b)\,P_{n}(\lambda',b)^{*} =\frac{\Gamma(2b)^{2}}{|\Gamma(b+i\lambda)|^{4}} 
&\int_{0}^{1}  \int_{0}^{1}ds\,dt\,[s(1-s)t(1-t)]^{b-1}\left(\frac{s}{s-1}\right)^{i\lambda}\left(\frac{t}{1-t} \right)^{-i\lambda'}Ê\nonumber \\
 &\left\{\sum_{n=0}^{\infty}[(2b)_{n}/n!]\,[(1-2s)(1-2t)]^{n}\right\}.
\end{eqnarray}
The sum under the integral signs can be evaluated: \[ \sum _{n=0}^{\infty}[(2b)_{n}/n!]\,[(1-2s)(1-2t)]^{n}=[1-(1-2s)(1-2t)]^{-2b} =[2(s+t-2st)]^{-2b}.\]
Making the variable substitutions $s=e^{x}/(1+e^{x}),\,t=e^{y}/(1+e^{y})$ we get for the double integral in (\ref{A10}), which we denote by $I$
\begin{equation} \label{A11}
I=2^{-2b}\int_{-\infty}^{\infty}\int_{-\infty}^{\infty}\left[\frac {e^{(x+y)/2}}{e^{x}+e^{y}}\right]^{2b}\,e^{i(\lambda x-\lambda'y)}\,dx\,dy.
\end{equation}

We can \lq\lq separate\rq\rq  the double integral by the substitution $x=u+v,\,y=-u+v$ to get
\begin{equation} \label{A12}
I=2^{-4b+1}\int_{-\infty}^{\infty}\frac{e^{i(\lambda+\lambda')u}}{(\cosh u)^{2b}} \,du \cdot \int_{-\infty}^{\infty}e^{i(\lambda-\lambda')v}\,dv=\frac{2^{-2b}|\Gamma(b+i\lambda)|^{2}}{\Gamma(2b)} 2\pi\,\delta(\lambda-\lambda'),
\end{equation}
where we have used the integral ([5],p.11,(26))
\begin{equation} \label{A13}
\int_{-\infty}^{\infty}\frac{e^{i2\lambda u}}{(\cosh u)^{2b}} \,du=2\int_{0}^{\infty}\frac{e^{i2\lambda u}}{(\cosh u)^{2b}} \,du=2^{2b-1}|\Gamma(b+i\lambda)|^{2}/\Gamma(2b), 
\end{equation}
and the representation
 \[ \int_{-\infty}^{\infty} e^{i(\lambda-\lambda')v}\,dv=2\pi\,\delta(\lambda-\lambda'), \]
 and furthermore used the $\delta$-function to put $\lambda'=\lambda$ in the first integral. 
 
 Putting (\ref{A10}), (\ref{A11}), and (\ref{A12}) together we have obtained the completeness relation
 \begin{equation} \label{A14}
 \sum_{n=0}^{\infty}P_{n}(\lambda,b)\,P_{n}(\lambda',b)=\delta(\lambda-\lambda')/\rho_{b}(\lambda).
 \end{equation}
 
 \subsection{Orthonormality of the set $\{P_{n}\}$ using an integral representation}
 
 We want to show thatÊ$\int_{-\infty}^{\infty}P_{m}(\lambda ,b)\,P_{n}(\lambda ,b) \rho_{b}(\lambda) \, d\lambda=\delta_{m n}$. For this purpose it is convenient to use the following contour integral representation of a $_{2}F_{1}$ function with the first parameter  equal to a non-positive integer -$n$, i.e. a polynomial of degree $n$.. The validity is established by expanding the factor $(1-tz)^{-b}$.
\begin{equation} \label{A15}
_{2}F_{1}(-n,b,c;z)=\frac{(-1)^{n} n!}{(c)_{n}}\frac{1}{2\pi i}\oint t^{-n-1}(1-t)^{c+n-1}(1-tz)^{-b}\,dt,
\end{equation}
where the contour only circles the singularity $t=0$.
Using (\ref{A15}) in the orthonormality integral and inverting integration orders we get
\begin{eqnarray} \label{A16}
I_{mn}=\int_{-\infty}^{\infty}P_{m}(\lambda,b)^{*}\,P_{n}(\lambda,b)\,\rho_{b}(\lambda)\,d\lambda =\sqrt{m!n!/(2b)_{m}(2b)_{n} }\,i^{m+n}(-1)^{n}2^{2b-1}/\pi\Gamma(2b) \times \nonumber \\ \frac{1}{(2\pi i)^{2}}\oint\oint ds\,dt\,\left\{\int_{-\infty}^{\infty}|\Gamma(b+i\lambda)|^{2} \,e^{i\lambda[\ln(1-2s)-\ln(1-2t)]}\,d\lambda \right\} \times \nonumber \\
\frac{s^{-m-1}t^{-n-1}(1-s)^{2b+m-1}(1-t)^{2b+n-1}} {[(1-2s)(1-2t)]^{b}}. 
\end{eqnarray}
The $\lambda$-integral in this formula is the Fourier inverse of the integral in (\ref{A13}); thus
\begin{equation} \label{A17}
\int_{-\infty}^{\infty}|\Gamma(b+i\lambda)|^{2}e^{i\lambda u}\,d\lambda =\pi 2^{-2b+1} \Gamma(2b)/(\cosh u/2)^{2b}.
\end{equation}
With $u=\ln[(1-2s)/(1-2t)]$  we get $(\cosh u/2)^{2}=(1-s-t)^{2}/(1-2s)(1-2t)$, and making the coordinate transformations $\sigma=s/(1-s),\,\tau=t/(1-t)$ we arrive at
\begin{equation} \label{A18}
I_{mn}=\sqrt{m!n!/(2b)_{m}(2b)_{n}}i^{m+n}(-1)^{n}\frac{1}{(2\pi i)^{2}} \oint\oint d\sigma\,d\tau\, \sigma^{-m-1}\tau^{-n-1}(1-\sigma\tau)^{-2b}.
\end{equation}
Expanding $(1-\sigma\tau)^{-2b}=\sum_{k=0}^{\infty}[(2b)_{k}/k!](\sigma\tau)^{k}$ in the last integral and performing the contour integrations we arrive at the desired result \[\int_{-\infty}^{\infty}P_{m}(\lambda ,b)\,P_{n}(\lambda 
,b) \rho_{b}(\lambda)\, d\lambda=\delta_{m n}.  \]

\subsection{Some comments on the first derivation of completeness and its relation to Jacobi matrices and the classical Hamburger moment problem}

It follows from the discussion in Section 2 , comparing (12), (17), (18), and (31), that the completeness relations for the Pollaczek polynomials are (for $b=1/4$ and $3/4$)  generalized orthonormality relations for the generalized eigenfunctions of the generator $J_{1}=(a^{\dagger 2}+a^{2})/4$. Correspondingly the set of recursion relations (\ref{A3}) is the eigenvalue equation  $Af=\lambda f$ for the infinite Jacobi matrix

\begin{equation} \label{A19}
  A= \left (
				\begin{array}{cccc} 
					0 & c_{0} & 0 & ...     \\
			    c_{0} & 0 &c_{1} & 0 \\
					 0 & c_{1} & 0 &c_{2}\\
					..  & 0 & c_{2}  & 0     
				\end{array}
				\right)  ,  f=
			\left  (
  \begin{array}{c}
		P_{0}\\
		P_{1}\\
		P_{2}\\
			..
		\end{array}
		\right  ); \mbox{ we have all } c_{m}>0
\end{equation}

Our case is a special kind of Jacobi matrix in the sense that the diagonal matrix elements of $A$ are all zero; for a general Jacobi matrix they can take any (real) values. The connection between the Jacobi matrix and the Hamburger moment problem is as follows:
The Hamburger moment problem is to find a (positive) measure $\tau$ on the real line corresponding to the moments $s_{m}=\int_{-\infty}^{\infty}u^{m}d\tau(u),\: m=0,1,2,..$. A necessary and sufficient condition for the existence of a solution is that the sequence $\{s_{m}\}$ is positive, i.e. that the Hankel forms $\sum_{i,k=0}^{n}s_{i+k}x_{i}x_{k}>0$ for non-zero vectors $ \{x_{0},x_{1},..\}$. This condition is equivalent to the requirement that a sequence of determinants of the Hankel forms should be positive. Certain combinations of the determinants determine the coefficients $c_{m}$ in a Jacobi matrix; expressed in properties of the associated Jacobi matrix the condition for the solvability of the corresponding moment problem is that all $c_{m}>0$.

The Jacobi matrix belongs to {\it type D} (limit point case; the corresponding moment problem is determined, i.e. there is a unique solution for the measure), provided  $\sum1/c_{m}=\infty $ (sufficient but not necessary condition!). Then the symmetric operator $A$, defined e.g. on finite sequences $f$, has a closure which is self-adjoint. It is easily seen that the condition on  $\{c_{m}\}$ is satisfied in our case. The unique measure is then supported by the spectrum of the self-adjoint operator, and is identical to our weight function $\rho_{b}$ The moments $s_{m}=\int_{-\infty}^{\infty}u^{m}d\tau(u)$ can then evidently be obtained from the Fourier transform of the weight function (\ref{A17}) as
\begin{equation} \label{A20}
\int_{-\infty}^{\infty}\lambda^{2m}\rho_{b}(\lambda)\,d\lambda=(-1)^{m}[d^{2m}/dx^{2m}(\cosh x/2)^{-2b}]_{x=0}
\end{equation}
The odd moments all vanish, corresponding to the fact that the diagonal Jacobi matrix elements are zero.

We refer to \cite{N97} for a more extensive discussion of the connection with Jacobi matrices and the Hamburger moment problem, in particular regarding the case of \lq\lq higher power squeezing generators\rq\rq $A_{k}=a^{k}+a^{\dagger k},\:k\geq 3$. In each of the subspaces $H_{\kappa}=$ linear span of $\{|k\,m+\kappa\rangle;\,m=0,1,..\}, \:\kappa=0,1,..,k-1$ which reduce the eigenvalue equation for $A_{k}$ the Jacobi matrix derived from the recursion relation corresponds to an undetermined moment problem. There is an infinity of solutions, and correspondingly the generator  $A_{k}$ is symmetric with deficiency indices $(1,1)$ in each subspace. There is in such a subspace a one-parameter family of self-adjoint extensions.

\end{document}